\documentclass[eqsecnum]{revtex4}
\begin{document}
\title{Critical analyses of order parameter and phase transitions at high density 
in Gross-Neveu model\footnote{ This work was partially supported by the National Natural Science Foundation of China.}
\\
}
\author{Bang-Rong Zhou \\
Department of Physics, Graduate School of the \\
Chinese Academy of Sciences, Beijing 100039, China and \\
CCAST(World Lab.), P.O. Box 8730, Beijing 100080, China }
\begin{abstract}
By critical analyses of the order parameter of symmetry breaking, we have researched 
the phase transitions at high density in $D=2$ and $D=3$ Gross-Neveu (GN) model and 
shown that the gap equation obeyed by the dynamical fermion mass has the same 
effectivenesss as the effective potentials for such analyses of all the second order 
and some special first order phase transitions. In the meantime we also further ironed 
out a theoretical divergence and proven that in $D=3$ GN model a first order phase 
transition does occur in the case of zero temperature and finite chemical potential.
\end{abstract}
\maketitle

\section{Introduction\label{sec:intro}}
The symmetry restoring phase transition at high temperature and high density has been 
an interesting research topic which is closely relative to evolution of early universe 
and phase transition of the nuclear matter \cite{kn:1,kn:2,kn:3,kn:4}. A good 
laboratory to research the phase transition is the models of four-fermion 
interactions including the Gross-Neveu(GN) models in $D=2$ and $D=3$ \cite{kn:5} 
time-space dimensions. On the phase transitions in these models, extensive 
investigations have been made \cite{kn:6,kn:7,kn:8,kn:9, kn:10, kn:11, kn:12, kn:13}, 
some common conclusions have been reached  but there is still particular question, 
e.g. whether first order phase transition exists or not in $D=3$ GN model, which needs 
to be further clarified. \\
\indent For determining the order of a phase transition, one can analyze critical 
behaviors of dynamical fermion mass as the order parameter of symmetry breaking at 
a critical temperature and/or at a critical chemical potential. The phase transition 
will be second or first order if the dynamical fermion mass goes to zero at the critical 
point continuously or discontinuously. \\
\indent The fundamental approach of the critical analyses should be using an effective 
potential. However, in many cases, the gap equation is a more simple and direct means. 
So called the gap equation we mean the essential part of the Schwinger-Dyson equation 
obeyed  by the dynamical fermion mass as the order parameter of symmetry breaking. 
It can be independently derived from the fermion condensates induced by the 
four-fermion interactions without using an effective potential, though it has 
been proven that the gap equation can also come from the extreme value condition of 
an effective potential \cite{kn:14}. If a phase transition is second order and this 
normally happens when the gap equation has the only non-zero solution, then the 
critical analysis can be made simply by means of gap equation. This has been carried 
out in the demonstrations on high temperature second order phase transition for fixed 
chemical potential in $D=4$ Nambu-Jona-Lasinio model \cite{kn:15},$D=3$ \cite{kn:16} and $D=2$ \cite{kn:17} GN model. However, if a phase transition is first order and 
this is usually the case when the gap equation has two or more non-zero solutions, 
then how should one make the critical analysis ?  Whether or in what conditions can 
one do that simply based on the gap equation ?  \\
\indent To answer the above questions, in this paper we will generalize 
critical analyses to the case including first order phase transitions. As far as the 
four-fermion models are concerned, it is most possible that the first order phase 
transition occurs in the case of high density. Hence we will investigate symmetry 
restoring phase transitions at high density for a given temperature, especially for 
zero temperature case, in $D=2$ and $D=3$ GN models. We will use both gap equations 
and effective potentials, and specially examine effectiveness and limitations of the 
gap equation for critical analyses of first order phase transition. \\
\indent The Lagrangian of the models will be expressed by
\begin{equation}
{\cal L}(x)=\sum_{k=1}^N\bar{\psi}^k(x)i\gamma^{\mu}\partial_{\mu}\psi_k(x)+
              \frac{g}{2}\sum_{k=1}^N[\bar{\psi}^k(x)\psi_k(x)]^2,
\end{equation}
where $\psi_k(x)$  are the fermion fields with $N$ "color" components and $g$ is 
the four-fermion coupling constant. The discussions will be made in the fermion bubble 
diagram approximation which is equivalent to the leading order of the $1/N$ expansion. 
It is assumed that higher-order corrections in $1/N$ do not change the qualitative 
feature of the models, since our models involve only some discrete symmetries which 
will be spontaneously broken or restored \cite{kn:18}. Our basic starting point is 
the gap equations coming from the fermionic bilinear condensates induced by the 
four-fermion interactions in Eq.(1.1), besides the corresponding effective 
potentials. The gap equations have been given in Ref.\cite{kn:16,kn:17}.  The 
corresponding effective potentials may be obtained from the relations between the 
gap equations and the effective potentials proven in Ref.\cite{kn:14}. Here we will 
directly use the known explicit results of them. \\
\indent The paper is arranged as follows. In Sect.~\ref{sec:2D} we discuss $D=2$ GN 
model, including analyzing critically the second order high density phase transition 
at high temperature $T$ by the gap equation and in particular, the first order high density phase transition at $T=0$ by the gap equation combined with the corresponding 
effective potential.  In Sect. ~\ref{sec:3D}, similar discussions to ones 
in Sect.~\ref{sec:2D} will be made in $D=3$ GN model. The first order high density 
phase transition at $T=0$ in this model will be proven respectively by critical 
analyses based on the gap equation and the effective potential. A difference between 
it and general first order phase transition and this difference's significance for 
usefulness of the gap equation analysis will be indicated. Finally in 
Sect.~\ref{sec: conclusions} we come our conclusions. 
\section{$D=2$ Gross-Neveu model\label{sec:2D}}
\indent In this model\cite{kn:17}, there exists a matrix 
$\gamma_5=-\sigma^3$ anticommunicating with $\gamma^0=\sigma^1$ and 
$\gamma^1=i\sigma^2$, where $\sigma^i(i=1,2,3)$ are the Pauli matrices. By the 
Mermin-Wagner-Coleman theorem, a continuous symmetry cannot be spontaneously broken 
in 2 dimensions \cite{kn:19}. Dynamical generation of the fermion mass will 
spontaneously break only the discrete chiral symmetry 
$\chi_D: \psi (t,x)\stackrel{\chi_D}{\longrightarrow} \gamma_5\psi 
(t,x)$ and  the special parity ${\cal P}_1 : \psi (t,x)\stackrel{{\cal 
P}_1}{\longrightarrow} \gamma^1\psi (t,-x)$. Hence possible phase transition at
high temperature and high density will relate to restoration of only the $\chi_D$
and ${\cal P}_1$. However, in 2 dimensions, symmetry restoration could occur only 
in the mean-field approximation \cite{kn:6,kn:20} which is equivalent to fermion 
bubble diagram approximation \cite{kn:17}.Because the former can be used only if 
$N\to \infty$ or $N$ is large and only a finite segment of one-space dimension
system is considered \cite{kn:20}, here it is assumed that the use of the fermion 
bubble diagram approximation will be in the above conditions. In 2 dimensions, 
Lagrangian (1.1) is perturbatively renormalizable. Assume the four-fermion 
interactions in Eq.(1.1) could lead to the condensates  
$\sum_{k=1}^N\langle\bar{\psi}^k\psi_k\rangle\neq 0$ when $T=\mu=0$ and the thermal 
condensates $\sum_{k=1}^N{\langle\bar{\psi}^k\psi_k\rangle}_T\neq 0$ when $T\neq 
0$, where $\mu$ is the chemical potential of fermions, then we will obtain the 
renormalized gap equations in respective cases and finally reduce the gap equation 
at $T\neq 0$ to the following form
\begin{equation}
\ln \frac{m(0)}{m}=F_1(T,\mu,m)=I_1(y,-r)+I_1(y,r)
\end{equation}
with the denotations 
\begin{equation}
I_1(y,\mp r)=\int_{0}^{\infty}\frac{dx}{\sqrt{x^2+y^2}}
\frac{1}{\exp(\sqrt{x^2+y^2}\mp r)+1},
\end{equation}
where $y=m/T$, $r=\mu/T$, $m(0)$ and $m\equiv m(T,\mu)$ are respectively the 
dynamical fermion masses at $T=\mu=0$ and at $T\neq 0$ and/or $\mu\neq 0$.  It is 
easy to verify that $F_1(T,\mu, m)$ is an increasing function of $T$ and $\mu$ but 
a decreasing function of $m$, and from  Eq.(2.1), we can assume that at  some critical 
temperature $T_c$ and critical chemical potential $\mu_c$, the dynamical fermion mass 
$m$ becomes zero. This will lead to a critical equation. At a high temperature $T$, 
by using the high temperature expansion of $F_1(T,\mu,m)$ \cite{kn:15} we may reduce 
the gap equation (2.1) to 
\begin{equation}
\matrix{\ln\frac{m(0)}{T\pi}+\gamma=g(T,\mu)(\frac{y}{2\pi})^2
+h(T,\mu)+ {\cal O}\left[(\frac{y}{2\pi})^4\right], \vspace{0.3cm}\cr
g(T,\mu)=\frac{7}{2}\zeta (3)-93\zeta (5)(\frac{r}{2\pi})^2
+\frac{1905}{2}\zeta (7)(\frac{r}{2\pi})^4, \vspace{0.3cm}\cr
h(T,\mu)=7\zeta(3) (\frac{r}{2\pi})^2-31\zeta(5) (\frac{r}{2\pi})^4
+127\zeta (7)(\frac{r}{2\pi})^6, \vspace{0.3cm}\cr}
\end{equation}
where $\zeta(s)(s=3,5,7)$ are the Riemann zeta functions and $\gamma$ is the Euler
constant. Noting that the $\ln m$ term in Eq.(2.1) has been cancelled, thus we can 
set $y=0$ in Eq.(2.3) and obtain the critical equation
\begin{equation}
\ln\frac{m(0)}{T_c\pi}+\gamma= h(T_c,\mu_c).
\end{equation}
In the derivation of Eq. (2.4) an implicit fact is that  Eq. (2.3) has the only 
solution $m$ and $m$ can go to zero continuously as $T$ and $\mu$ vary. This is 
just the characteristic of a second order phase transition. In view of this, Eq. (2.4) 
can be called the critical equation of a second order phase transition. The critical 
points determined by Eq. (2.4) are all second order ones. For verifying this 
conclusion, we can discuss from Eqs.(2.3) and (2.4) critical behaviors of the squared 
order parameter $m^2$ at $\mu_c$ when $T$ is fixed. Replacing $T_c$ by $T$ in Eq.(2.4) 
and substituting it into Eq. (2.3), we get
\begin{equation}
m^2\simeq (\mu_c^2-\mu^2) f(T,\mu_c)/g(T,\mu),\ 
\ {\rm when} \ \mu \sim \mu_c,
\end{equation}
where 
\begin{equation}
f(T,\mu_c)= 7\zeta (3)-62\zeta (5)(\frac{\mu_c}{2\pi T})^2
+381\zeta (7)(\frac{\mu_c}{2\pi T})^4.
\end{equation}
Eq.(2.5) shows that the squared order parameter $m^2$ goes to zero continuously when 
$\mu\to \mu_c$ in the form of $\mu_c-\mu$, hence for a given high temperature $T$, 
the symmetry restoration phase transition at $\mu_c$ is second order indeed.\\
\indent Eq.(2.4) comes from high temperature expansion of $ F_1(T,\mu,m)$ and is 
 inapplicable to discussion of high density phase transition at low temperature and 
$T=0$. For discussion of the latter, we have to use the effective potential. In the following we will only focus on the high density phase transition at 
$T=0$. Let us first go back to Eq.(2.1). Through changing the integral variable 
by $z=(x^2/y^2+1)^{1/2}$ we can express 
\begin{equation}
F_1(T,\mu,m)=\int_1^{\infty}\frac{dz}{\sqrt{z^2-1}}\left[\frac{1}{e^{y(z-\alpha)}+1}
+(-\alpha \to \alpha)\right],
\end{equation}
where $\alpha=\frac{\mu}{m}$. In the limit $T\to 0$ or $y\to \infty$, we have 
\begin{equation}
F_1(T=0,\mu,m)=\theta(\mu-m)\ln\frac{\mu+\sqrt{\mu^2-m^2}}{m}.
\end{equation}
Substituting Eq.(2.8) into Eq.(2.1) we obtain
\begin{equation}
\ln\frac{m(0)}{m}=\theta(\mu-m)\ln\frac{\mu+\sqrt{\mu^2-m^2}}{m}
\end{equation}
which has the solutions
\begin{equation}
m=m(0), \   {\rm when} \ \mu\leq m \\
\end{equation}
and
\begin{eqnarray}
m&=&\{m(0)[2\mu-m(0)]\}^{1/2}\equiv m_1, \nonumber \\ 
&&{\rm when} \ \mu>m \ {\rm and} \ \frac{m(0)}{2}\leq \mu < m(0).
\end{eqnarray}
On the other hand, the effective potential $V_{eff}^{(2)}(T,\mu,m)$ has been proven 
to satisfy the equation \cite{kn:14}
\begin{equation}
\frac{\partial V_{eff}^{(2)}(T,\mu,m)}{\partial m}
=\frac{m}{\pi}\left[\ln\frac{m}{m(0)}+F_1(T,\mu,m)\right],
\end{equation}
noting that the gap equation (2.1) multiplied by $m$ is just the extreme conditions 
given by $\partial V_{eff}^{(2)}(T,\mu,m)/\partial m =0$. Hence, when $T=0$, 
$V_{eff}^{(2)}(T=0,\mu,m)$ for  $m(0)/2\leq \mu < m(0)$ will have the three possible extreme points $m=0,$ $m_1$ and $m(0)$. To determine which one of them corresponds to a minimum or a maximum of $ V_{eff}^{(2)}(T=0,\mu,m)$, we may calculate  and  obtain directly from Eq. (2.12)
\begin{equation}
\frac{\partial^2V_{eff}^{(2)}(T=0,\mu,m)}{\partial m^2}=\frac{1}{\pi}
\left\{\matrix{
\ln\frac{m}{m(0)}+1          &{\rm when} &\mu\leq m \cr
\ln\frac{\mu+\sqrt{\mu^2-m^2}}{m(0)}-\frac{m^2}{(\mu+\sqrt{\mu^2-m^2})\sqrt{\mu^2-m^2}} &{\rm when } &\mu>m \cr}\right.,
\end{equation}
where Eq. (2.8) has been used. Eq. (2.13), combined with Eqs. (2.10) and (2.11),
give that 
\begin{equation}
\left.\frac{\partial^2V_{eff}^{(2)}(T=0,\mu,m)}{\partial m^2}\right|_{m=0}=
\frac{1}{\pi}\ln\frac{2\mu}{m(0)}\left\{\matrix{<0 &{\rm when}&\mu<m(0)/2\cr
                              >0 &{\rm when}&\mu>m(0)/2\cr}\right.,
\end{equation}
\begin{equation}
\left.\frac{\partial^2V_{eff}^{(2)}(T=0,\mu,m)}{\partial m^2}\right|_{m=m_1}=-\frac{1}{\pi}\frac{2\mu-m(0)}{m(0)-\mu}<0,\ 
{\rm when} \ \frac{m(0)}{2}<\mu<m(0)
\end{equation}
and
\begin{equation}
\left.\frac{\partial^2V_{eff}^{(2)}(T=0,\mu,m)}{\partial m^2}\right|_{m=m(0)}=\frac{1}{\pi}>0, \ {\rm when} \ \mu<m(0).
\end{equation}
Eqs. (2.14)-(2.16) indicate that \\
1) When $\mu<\frac{m(0)}{2}$, $ V_{eff}^{(2)}(T=0,\mu,m)$ will have the 
maximum point $m=0$ and the minimum point $m=m(0)$, thus the model has the same 
spontaneous symmetry breaking  as one at $\mu=0$.\\
2) When $\frac{m(0)}{2}<\mu<m(0)$, $ V_{eff}^{(2)}(T=0,\mu,m)$ will have three 
extreme points: $m=0$ (minimum),$m=m_1$ (maximum) and $m=m(0)$ (minimum), and this 
is just the characteristic of possible emergence of a first order phase transition. 
\\
\indent It is emphasized that in the above analyses we only use the known relation 
(2.12) between the gap equation and the effective potential and the explicit 
expression of the gap equation (2.1), and did not use an explicit expression of 
$ V_{eff}^{(2)}(T,\mu,m)$.  However, for determining the critical point, we have to 
find out the explicit expression of $ V_{eff}^{(2)}(T,\mu,m)$ from Eq.(2.12).
The result is \cite{kn:14}
\begin{eqnarray}
V_{eff}^{(2)}(T=0,\mu,m)&=&\frac{m^2}{2\pi}\left[\ln\frac{m}{m(0)}-\frac{1}{2}\right]+\frac{\mu^2}{2\pi} +\frac{1}{2\pi}  \theta(\mu-m)\nonumber \\ 
&&\times \left(m^2\ln\frac{\mu+\sqrt{\mu^2-m^2}}{m}-\mu\sqrt{\mu^2-m^2}\right).
\end{eqnarray}
Assuming that $\frac{m(0)}{2}<\mu<m(0)$, then for $m=m(0)$ among the three extreme 
points, we will get
\[
V_{eff}^{(2)}[T=0,\mu,m=m(0)]=\frac{1}{4\pi}[2\mu^2-m^2(0)]<0,
\]
\begin{equation}
{\rm when} \ \mu<\frac{1}{\sqrt{2}}m(0).
\end{equation}
Hence, if $\mu<m(0)/\sqrt{2}$, $m=m(0)$ will be the global minimum point of 
$V_{eff}^{(2)}(T=0,\mu,m)$, because in this case, the minimum 
$V_{eff}^{(2)}(T=0,\mu,m=0)=0$ and $V_{eff}^{(2)}(T=0,\mu,m=m_1)$ is a maximum. 
Obviously, the first order phase transition point should be determined by
\begin{equation}
V_{eff}^{(2)}[T=0,\mu,m=m(0)]= V_{eff}^{(2)}(T=0,\mu,m=0)=0
\end{equation}
and this will lead to
\begin{equation}
\mu=\frac{1}{\sqrt{2}}m(0)\equiv \mu_c,
\end{equation}
i.e. when $\mu$ crosses over $\mu_c$, the global minimum point of 
$V_{eff}^{(2)}(T=0,\mu,m)$ will jump from $m(0)$ down to 0 through a barrier 
tunneling between the minimum points $m=m(0)$ and $m=0$. The critical chemical 
potential $\mu_c$ of the first order phase transition at $T=0$ in $D=2$ GN model in 
the leading order of $1/N$ expansion given by Eq. (2.20) is consistent with  the one 
obtained by other authors \cite{kn:7,kn:8, kn:9, kn:10, kn:11, kn:12}. Since the 
global minimum of $V_{eff}^{(2)}(T=0,\mu,m)$ is also at $m=m(0)$ when $\mu<m(0)/2$, 
we can write the critical behavior of the order parameter $m$ by 
\begin{equation}
m=\left\{\matrix{m(0) &{\rm when} &\mu\leq \frac{1}{\sqrt{2}}m(0)\cr
                 0    &{\rm when} & \mu>\frac{1}{\sqrt{2}}m(0)\cr}\right..
\end{equation}
The above discussions show that for critical analyses of the order parameter
in a general first order phase transition, it is not sufficient only using the gap
equation, we must use the gap equation combined with corresponding effective potential.  This is also applicable to critical analyses of  the phase transitions at low temperature $T\neq 0$ in $D=2$ GN model which are known to be first order.  The effective potential at $T\neq 0$ may be derived from Eq.(2.12). By means of it and the gap equation (2.1),  we will be able to obtain  other results of the phase transition at low temperature in this model including location of the tricritical point etc., following similar methods to ones used in Ref \cite{kn:6,kn:7,kn:11,kn:12}.
\section{$D=3$ Gross-Neveu model\label{sec:3D}}
\indent  In this case \cite{kn:16}, the dynamical fermion 
mass will spontaneously break the special parities ${\cal P}_1$ and ${\cal P}_2$: 
$\psi(t, x^1, x^2)$ $\stackrel{{\cal P}_1}{\longrightarrow}\gamma^1\psi(t,-x^1, 
x^2)$ and $\psi(t, x^1, x^2)\stackrel{{\cal P}_2}{\longrightarrow}
\gamma^2\psi(t,x^1, -x^2)$ and the time reversal ${\cal T}$: 
$\psi(t, x^1, x^2)\stackrel{{\cal T}}{\longrightarrow}
\gamma^2\psi(-t,x^1, x^2)$\footnote{We no longer include the discrete symmetry $Z_2$: $\psi^1_k(x)\stackrel{Z_2}{\longleftrightarrow}\psi^2_k(x), \ \ {\rm if} \
\psi_k(x)=\left(\psi^1_k(x),\psi^2_k(x) \right)$, as was stated in 
Ref.\cite{kn:16}, since the kinetic energy terms in Eq. (1.1) actually have not 
separate $Z_2$ symmetry.}. After the gap equation at $T=\mu=0$ is substituted,
the gap equation at $T\neq 0$ becomes 
\begin{equation}
m(0)=m+F_2(T,\mu,m)
\end{equation}
with
\begin{equation}
F_2(T,\mu,m)= T\left\{\ln \left[1+e^{-(m-\mu)/T}\right]+
                  (-\mu\to \mu)\right\},
\end{equation}
and the critical equation of second order phase transition will be
\begin{equation}
m(0)= T_c\left[\ln (1+e^{\mu_c/T_c})+
                  \ln (1+e^{-\mu_c/T_c})\right]
\end{equation}
which determines critical points of second order phase transitions.  For example, for a given finite $T$, when $\mu\sim \mu_c$, $m\approx 0$, so we will 
have $m/T \ll 1$.  Then  Eq.(3.1) with (3.2) will lead to 
\begin{eqnarray}
m(0)&=&T\ln\left(1+e^{\mu/T}\right)+T\ln\left(1+e^{-\mu/T}\right)\nonumber \\
&&+\frac{m^2}{2T[1+\cosh(\mu/T)]}+{\cal O}\left(\frac{m^3}{T^3}\right).
\end{eqnarray}
Substituting Eq.(3.3) with $T_c$ replaced by $T$ into Eq.(3.4) we get
\begin{equation}
m^2=2T\sinh(\mu/T)(\mu_c-\mu), \ {\rm for} \ T\neq 0.
\end{equation}
Eq.(3.5) shows that for a given finite temperature, the symmetry restoration phase 
transition at $\mu_c$ is second order. The same has been shown for the phase transition 
at $T_c$ for a given $\mu$ \cite{kn:16}. These conclusions are valid for $T\neq 0$. 
For $T=0$, we must go back the gap equation (3.1). We note that when $\mu\leq m$, 
$\lim \limits_{T\to 0}F_2(T,\mu,m)=0$ and by Eq.(3.1) this implies that $m=m(0),\ 
\ {\rm if} \ \ \mu\leq m$, or furthermore, $ m=m(0),\ \ {\rm if} \ \ \mu\leq m(0)$.  
On the other hand, if $\mu>m$, then  $\lim \limits_{T\to 0}F_2(T,\mu,m)=\mu-m$. 
Substituting this result into Eq.(3.1) we will obtain that $m(0)=\mu$, but the 
presupposition leading to this result in fact should be $\mu>m=m(0)$, the both 
apparently contradict each other. This indicates  that when $\mu>m=m(0)$, the gap 
equation (3.1) can not be satisfied, thus there is no longer symmetry breaking and 
the order parameter $m$ must be zero.  The above results can be summarized as 
\begin{equation}
m=\left\{\matrix{m(0) & {\rm when} &\mu\leq m(0)\cr
                 0    & {\rm when} &\mu> m(0)   \cr}\right..
\end{equation}
Furthermore, when $T\to 0$, the derivative $\partial m/\partial \mu$ will be
\begin{eqnarray}
&&\left.\lim \limits_{T\to 0}\frac{\partial m}{\partial \mu}\right|_{m=m(0)}
\nonumber \\
&&=\left.-\lim\limits_{T\to 0}\frac{\partial F_2(T,\mu,m)}{\partial \mu}/
\left[1+\frac{\partial F_2(T,\mu,m)}{\partial m}\right]\right|_{m=m(0)}
\nonumber \\
&&=-\lim \limits_{T\to 0}e^{-[m(0)-\mu]/T}=\left\{\matrix{0, &{\rm when} \ \mu<m(0) \cr
-1, & {\rm when} \ \mu=m(0) \cr
-\infty, & {\rm when} \ \mu\agt m(0) \cr}\right.\nonumber \\
\end{eqnarray}
(noting that when $\mu>m(0)$, the gap equation (3.1) is no longer valid). 
Eqs.(3.6) and (3.7) show that when $\mu<m(0)$, $m$ does not change as $\mu$ varies 
and keeps its value $m(0)$ at $\mu=0$; however as soon as $\mu$ crosses over $m(0)$,
$\partial m/\partial \mu$ will change from 0 into $-\infty$ and $m$ will jump 
discontinuously from $m(0)$ to zero. This implies that when $T=0$ the phase transition 
at the critical chemical potential $\mu_c=m(0)$ is first order, i.e. 
$(T,\mu)=(0,m(0))$ will be a first order phase transition point in $T-\mu$ phase 
diagram. On the other hand, we note that $(T,\mu)=(0,m(0))$ is a solution of the 
critical equation (3.3) corresponding to second order phase transition when we take 
the limit $T\to 0$ in Eq.(3.3). Hence we may reasonably conclude that 
$(T,\mu)=(0,m(0))$ will be a tricritical point in the $T-\mu$ phase giagram of $D=3$ 
GN model. \\
\indent This conclusion coincides with the one in Ref. \cite{kn:13} but 
different from the one in Ref. \cite{kn:12} where the authors claimed that there is 
no first order phase transition in $D=3$ GN model. To clarify this problem, we can 
make further analysis of the phase transition at $T=0$ in the model by means of the 
effective potential which may be derived from the gap equation, as was made in $D=2$ 
GN model.  It has been proven that the extreme value condition of the effective 
potential $ V_{eff}^{(3)}(T,\mu,m)$ is \cite{kn:14}
\begin{equation}
\frac{\partial V_{eff}^{(3)}(T,\mu,m)}{\partial m}=0
\end{equation}
with
\begin{equation}
\frac{\partial V_{eff}^{(3)}(T,\mu,m)}{\partial m}=
\frac{m}{2\pi}\left[m-m(0)+F_2(T,\mu,m)\right],
\end{equation}
noting that Eq.(3.8) is just the gap equation (3.1) multiplied by $m$.  From Eq. (3.9) 
we may obtain $V_{eff}^{(3)}(T,\mu,m)$ satisfying the condition 
$V_{eff}^{(3)}(T,\mu,m=0)=0$ expressed by
\begin{equation}
V_{eff}^{(3)}(T=0,\mu,m)=\frac{1}{2\pi}\left\{\frac{m^2}{2}[\mu\theta(\mu-m)-m(0)]+\theta(m-\mu)\left(\frac{m^3}{3}+\frac{\mu^3}{6}\right)\right\}.
\end{equation}
For fixed $\mu$, from Eq.(3.10) or the $T\to 0$ limit of Eq.(3.9),we find out
\begin{equation}
\frac{\partial V_{eff}^{(3)}(T=0,\mu,m)}{\partial m}=\frac{m}{2\pi}\{
[\mu-m(0)]\theta(\mu-m)+[m-m(0)]\theta(m-\mu)\}
\end{equation}
and
\begin{equation}
\frac{\partial^2 V_{eff}^{(3)}(T=0,\mu,m)}{\partial m^2}=\frac{1}{2\pi}\{
[\mu-m(0)]\theta(\mu-m)+[2m-m(0)]\theta(m-\mu)\}.
\end{equation}
Variation of $V_{eff}^{(3)}(T=0,\mu,m)$ as $\mu$ increases may be discussed by Eqs. (3.10)-(3.12). \\
1) $\mu<m(0)$. In this case, $V_{eff}^{(3)}[T=0,\mu<m(0),m]$ will 
have the maximum point $m=0 (m<\mu)$ and the minimum point $m=m(0) (m>\mu)$. It is 
indicated that location of the minimum point keeps at $m=m(0)$ which does not change 
as $\mu$ increases. However, the minimum of the effective potential at $m=m(0)$ 
\begin{equation}
V_{eff}^{(3)}[T=0,\mu<m(0),m]|_{m=m(0)}=\frac{1}{12\pi}[\mu^3-m^3(0)]
\end{equation}
will rise as $\mu$ increases. Nevertheless, these results only show the same spontaneous symmetry breaking as one at $T=\mu=0$.\\
2) $\mu=m(0)$. In this case, the effective potential will take the following form
\[
V_{eff}^{(3)}[T=0,\mu=m(0),m]=\frac{1}{2\pi}\theta[m-m(0)]
\]
\begin{equation}
\times\left[\frac{m^3(0)}{6}-\frac{m^2}{2}m(0)+\frac{m^3}{3}\right]
\end{equation}
which indicates that in the total real axis segment $0\leq m\leq m(0)$,
$V_{eff}^{(3)}[T=0,\mu=m(0),m]=0$, thus the total segment $0\leq m\leq m(0)$ 
including the original $m=m(0)$ when $\mu<m(0)$, will be the minimum points of
$V_{eff}^{(3)}[T=0,\mu=m(0),m]$. This situation corresponds to that when 
$\mu=m(0)$, the $T\to 0$ limit of the gap equation (3.1) has infinite non-zero 
solutions.\\
3) $\mu>m(0)$. In this case, $\partial V_{eff}^{(3)}[T=0,\mu>m(0),m]/\partial m=0$
 will have the only solution $m=0$ and by Eq. (3.12),
\begin{equation}
\left.\frac{\partial^2 V_{eff}^{(3)}[T=0,\mu>m(0),m]}{\partial m^2}\right|_{m=0}
=\frac{1}{2\pi}[\mu-m(0)]>0.
\end{equation}
Thus $m=0$ will be the only minimum point of $V_{eff}^{(3)}[T=0,\mu>m(0),m]$ and this 
implies that the broken symmetries will be restored. The critical chemical potential 
$\mu_c$ should be determined by the condition
\begin{equation}
\left.\frac{\partial^2 V_{eff}^{(3)}(T=0,\mu,m)}{\partial m^2}\right|_{m=0}=0,
\end{equation}
i.e. at $\mu_c$, $m=0$ must change from being a maximum point into being a minimum point.  The result is that $\mu_c=m(0)$.\\
\indent The above discussions show that as $\mu$ increases and finally crosses over 
$\mu_c=m(0)$, the global minimum point of the effective potential 
$V_{eff}^{(3)}(T=0,\mu,m)$ will jump from $m(0)$ down to 0 thus Eq. (3.6) is 
reproduced. The phase transition is first order indeed. The analysis based on effective potential leads to the same conclusion as the one only based on the gap equation, however the latter is obviously more simple and direct in this model. We
also indicate that the first order phase transition at $T=0$ in $D=3$ GN model has 
a distinct feature, i.e. the transit from broken phase to symmetry phase does not 
undergo a barrier tunneling, since when $\mu$ is increasing, as we have seen above, 
no barrier emerges from between the minimum points $m=0$ and $m=m(0)$ of 
$V_{eff}^{(3)}(T=0,\mu,m)$. This is apparently different from general first order 
phase transition, e.g. the one in $D=2$ GN model discussed in Sect. \ref{sec:2D}. 
It is assumed that it is just this feature that makes one be able to determine the 
critical behavior of the order parameter in the first order phase transition in $D=3$ 
GN model merely by the gap equation approach. 
\section{Conclusions \label{sec: conclusions}}
In this paper, we have discussed symmetry restoring phase transitions at high density 
in $D=2$ and $D=3$ GN model by means of analyses of critical behaviors of the dynamical 
fermion mass as order parameter of symmetry breaking. We have used both gap 
equation and effective potential approach. In these approaches, the gap equations 
multiplied by the dynamical fermion mass are simply the Schwinger-Dyson equations 
obeyed by the dynamical fermion mass and the effective potentials will be derived 
from the gap equations based on the fact that a gap equation multiplied by the 
dynamical fermion mass can also come from the extreme value condition of corresponding 
effective potential. This shows that the Schwinger-Dyson equations obeyed by the 
 dynamical fermion mass in fact contain the essential elements of the phase structure 
of the discussed models. We have found that for second order phase transitions and 
some specific first order phase transitions e.g. the one at high density and $T=0$ 
in $D=3$ GN model in which the jumping of the order parameter does not correspond 
to a barrier tunneling in an effective potential, the gap equation analysis alone 
has the same effect as the effective potential one. However, for general first order 
phase transitions, e.g. the one  at high density and $T=0$ and low $T\neq 0$ in $D=2$ 
GN model, we must combine the gap equation with the corresponding effective potential 
so as to obtain correct conclusions. \\
\indent Our analyses also reproduce some of known essential results of the phase 
transitions in $D=2$ and $D=3$ GN model. They include that in $D=2$ GN model, the 
phase transitions at high density will be second order when $T$ is high and first 
order when $T\to 0$; in $D=3$ GN model, the phase transition at high density will be second order when $T$ is finite and first order when $T=0$ thus one can conclude that $(T,\mu)=(0,m(0))$ is a tricritical point. The latter further clarifies the theoretical divergence of that if first order phase transition exists in $D=3$ GN model.

\end{document}